\newcommand{\refe}[1]{(\ref{#1})}
\documentstyle[aps,epsf,twocolumn,psfig,array]{revtex}

\begin{document}

\twocolumn[\hsize\textwidth\columnwidth\hsize\csname@twocolumnfalse\endcsname

\title{Off-Diagonal Geometric Phases}

\author{Nicola Manini$^{a,\dag}$ and F. Pistolesi$^{b,*}$} 
 
\address{$^a$European Synchrotron Radiation Facility,
B.P. 220, F-38043 Grenoble Cedex, France\\
$^b$Institut Laue Langevin, B.P. 156, F-38042 Grenoble Cedex 9, France
}

\date{\today}
\maketitle 

\begin{abstract}
We investigate the adiabatic evolution of a set of 
non-degenerate eigenstates of a parameterized Hamiltonian.
%
Their relative phase change can be related to geometric measurable
quantities that extend the familiar concept of Berry phase 
to the evolution of more than one state.
We present several physical systems where these concepts can be 
applied, including an experiment on microwave cavities for which 
off-diagonal phases can be determined from published data.
\end{abstract}


] \narrowtext

\newcommand{\sv}{{\bf s}}

Consider the adiabatic evolution of a set of nondegenerate normalized
eigenstates $|\psi_i(\sv)\rangle$ of a parameterized Hamiltonian $H(\sv)$.
The idea that, with a suitable definition, the phase of the scalar product
$\langle \psi_j(\sv _1)|\psi_j(\sv _2)\rangle$ contains a geometric,
measurable contribution dates back to Pan\-charat\-nam's pioneering work
\cite{Pancharatnam}.
In particular, 
when $\sv_1=\sv_2$ and the state $|\psi_j(\sv)\rangle$ is transported
adiabatically along a closed loop, the existence of a nontrivial phase
factor was discovered and put on a firm basis by Berry \cite{Berry}.
Since then, considerable work has been devoted to
interpretation \cite{Simon83,Anandan87,Christian93,Resta94,manolopoulos},
generalization \cite{Aharonov87,Samuel88,Zak89,Simon93,Mukunda93,Rabei99},
and experimental
determination \cite{Delacretaz,Tycko87,Weinfurter90,Lauber94,Wagh98} of
these geometric phase factors.
Surprisingly, for $\sv _1\neq \sv _2$, the phase relation of $\langle
\psi_j(\sv_1) |\psi_k(\sv_2)\rangle$ between two {\em different}
eigenstates has not been equally well investigated so far
\cite{simon:note}.

This is even more surprising if one considers that, for some pair of
points $\sv_1$ and $\sv_2$, it may occur that
$|\psi_k(\sv_2)\rangle=e^{i\,\alpha}|\psi_j(\sv_1)\rangle$ 
($k\neq j$).
This implies that both scalar products $\langle \psi_j(\sv
_1)|\psi_j(\sv _2)\rangle$ and $\langle \psi_k(\sv _1)|\psi_k(\sv
_2)\rangle$ vanish, and, as well known, the usual Pancharatnam-Berry
phase on any path connecting $\sv _1$ to $\sv _2$ is undefined for the
states $k$ and $j$.
The only phase information left is thus contained in the 
cross scalar products
 $\langle \psi_j(\sv _1)|\psi_k(\sv _2)\rangle$.

In this Letter we determine the measurable and geometric phase factors
associated to the off-diagonal matrix elements $\langle \psi_j(\sv
_1)|\psi_k(\sv _2)\rangle$ of the operator describing the evolution along a
general open path in the parameter space that connect $\sv _1$ to $\sv _2$.
We find a set of independent off-diagonal phase factors 
that exhaust the geometrical phase information carried by the 
basis of eigenstates along the path.
Analogously to the familiar Berry phase, the values of these phases
depend on the presence of degeneracies of the
energy levels in the parameters space.
The formalism is then applied to an experiment on quantum
billiards \cite{Lauber94}, where the off-diagonal phase factors can be
extracted directly from published experimental data.

In order to introduce the off-diagonal geometric phases, it is
convenient to consider the usual definition of the geometric phase of
one normalized state $|\psi_{j}(\sv)\rangle$ in terms of parallel
transport \cite{Berry,Anandan87,Aharonov87,Samuel88}.  
Given any path
$\Gamma$ that joins $\sv_1$ to $\sv_2$, the state parallel-transported
along it is defined by:
\begin{equation}
	|\psi_j^\parallel(\sv_2)\rangle
	=
	\exp \left\{ - \!\!\int_{\Gamma} \!\! 
	d\sv\cdot 
	\langle \psi_j(\sv) 
	|\nabla_\sv \psi_j(\sv)\rangle
	\right\} \, |\psi_{j}(\sv_2)\rangle
	\, .
\end{equation}
This fixes the phase of the state along the path in the
unique way satisfying
$
\langle \psi_j^\parallel(\sv) | 
\psi_j^\parallel(\sv+{\mathbf \delta}) \rangle
= 1+ O(\delta^2)
$ for $\delta\rightarrow 0$, 
{\em i.e.} having maximal projection on the ``previous'' state.
The {\em geometric} phase factor is then defined simply in terms of
the scalar product along the parallel evolution:
\begin{equation}
	\gamma_j^{\Gamma}
	\equiv
	 \Phi\!\left(U^{\Gamma}_{jj}\right)
	=
	\Phi\!\left( \langle \psi_{j}^{\parallel}(\sv_1) 
	|\psi_j^\parallel(\sv_2)\rangle \right)
	 \, ,
	\label{berrydef:eq}
\end{equation}
where $\Phi(z)=z/|z|$ for complex $z \neq 0 $.
$\gamma_j^{\Gamma}$ is
univocally determined by the sequence $\Gamma_j$ of states 
$|\psi_j(\sv)\rangle$, with $\sv$ varying along $\Gamma$.
Indeed, $\gamma_j^{\Gamma}$ is unchanged by a local ``gauge''
transformation:
\begin{equation}
	| \psi_j(\sv )\rangle 
	\rightarrow
	|\psi_j(\sv )\rangle 
	\; \exp[i \varphi_j(\sv)] 
	\label{gchange:eq}
\end{equation}
and by any reparametrization of the sequence of states $\Gamma_j$.
It is thus a geometric, measurable quantity.

In a similar way, we define \cite{manolopoulos:note}
the phase factors associated to the
off-diagonal elements of the parallel-evolution operator $U^{\Gamma}$:
\begin{equation}
	\sigma_{jk}^{\Gamma}
	\equiv
	 \Phi\!\left(U^{\Gamma}_{jk}\right)
	= 
	\Phi\!\left(\langle \psi_{j}^\parallel(\sv_1) |
		\psi_{k}^\parallel(\sv_2)\rangle \right)
	\,
	.
	\label{sigmadef:eq}
\end{equation}
Like $\gamma_j^{\Gamma}$, the phase factor $\sigma_{jk}^{\Gamma}$ is
independent of the path parametrization.
However, $\sigma_{jk}^{\Gamma}$ depends on the relative phase of the two
vectors $|\psi_j \rangle$ and $|\psi_k \rangle $ at $\sv_1$.
Indeed, under the gauge transformation
\refe{gchange:eq}, $\sigma_{jk}^{\Gamma}$ transforms as follows:
\begin{equation}	
	\sigma_{jk}^{\Gamma} 
	\rightarrow 
	\sigma_{jk}^{\Gamma} \, 
	\exp i [\varphi_{k}(\sv _1) - \varphi_{j}(\sv _1)] \,.
	\label{sigmagauge:eq}
\end{equation}
This shows that $\sigma_{jk}^{\Gamma}$ is arbitrary, thus non-measurable.
In order to define a gauge-invariant quantity, we
combine two $\sigma$'s in the following product:
\begin{equation}
	\gamma_{jk}^\Gamma=\sigma_{jk}^\Gamma \; \sigma_{kj}^\Gamma
	\, .
	\label{gammadef1:eq}
\end{equation}
This new phase factor $\gamma_{jk}^\Gamma$
is determined uniquely by the trajectories $\Gamma_j$ and $\Gamma_k$
of $|\psi_{j}\rangle$ and $|\psi_{k}\rangle$ in the Hilbert space.
The finding of the measurable geometric quantity
$\gamma_{jk}^\Gamma$ is the central result of this Letter.

\begin{figure}[tbh]
\centerline{ 
\psfig{file=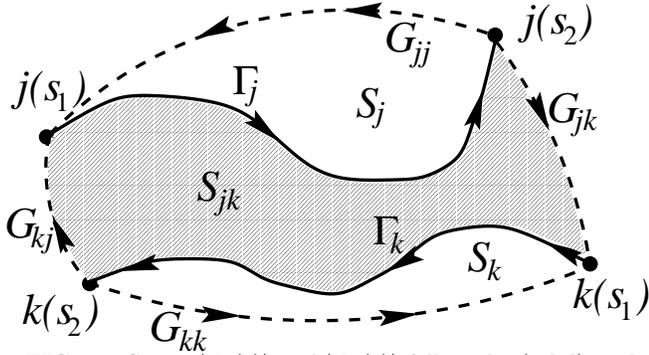,width=8.6cm}
}
\caption{
States $|\psi_j(\sv)\rangle$ and $|\psi_k(\sv)\rangle$ follow the
(solid) paths $\Gamma_j$ and $\Gamma_k$ along the evolution in rays space.
Geodesics $G_{jj}$, $G_{kk}$, $G_{jk}$, and $G_{kj}$ (dashed) lead back from
the evolved states $|\psi_j(\sv_2)\rangle$ $|\psi_k(\sv_2)\rangle$ to the
initial ones $|\psi_j(\sv_1)\rangle$ $|\psi_k(\sv_1)\rangle$.
Integration of Berry's 2-form over the shaded surface $S_{jk}$ yields the
off-diagonal phase $\gamma^\Gamma_{jk}$.
\label{surface:fig}
 }
\end{figure}

A simple geometric interpretation for $\gamma^\Gamma_{jk}$
can be obtained in analogy with that for the Pancharatnam 
phase.
Consider the path of state $j$ in the  space of rays
(where two states differing only for a complex factor are identified).
If $|\psi_j(\sv_1)\rangle$ is not orthogonal to $|\psi_j(\sv_2)\rangle$,
there exists a unique geodesic path $G_{jj}$ going from
$|\psi_j(\sv_2)\rangle$ to $|\psi_j(\sv_1)\rangle$, along which the
geometric phase factor is unity.
Then, trivially, the open-path geometric factor $\gamma_j^\Gamma$ equals
the phase factor on the circuit composed by $\Gamma_j$ and $G_{jj}$ (see
Fig.~\ref{surface:fig})\cite{Samuel88,Simon93}.
Once reduced to a closed path, using Stokes' theorem, one can write
$\gamma_j^\Gamma$ in terms of the integral of Berry's local-gauge-invariant
2-form on any surface $S_j$ bounded by $\Gamma_j+G_{jj}$
\cite{Berry,Samuel88,Mukunda93}.

Consider now {\em two} states $j$ and $k$ evolving along $\Gamma_j$ and
$\Gamma_k$ in the space of rays.
We generate all possible oriented loops by connecting
the extremal points  with geodesics.
As Fig.~\ref{surface:fig} shows, only the three loops $\Gamma_j+G_{jj}$,
$\Gamma_k+G_{kk}$ and $\Gamma_j+G_{jk}+\Gamma_k+G_{kj}$ can be generated.
The first two loops give the usual phase factors $\gamma^\Gamma_j$ and
$\gamma^\Gamma_k$, while the third one corresponds to $\gamma^\Gamma_{jk}$.
In this way, $\gamma^\Gamma_{jk}$ can be calculated, in analogy to
$\gamma^\Gamma_j$, as the integral of Berry's 2-form over a surface
$S_{jk}$ bounded by this 4-legs loop.
The complementarity of $\gamma^\Gamma_{jk}$ and $\gamma^\Gamma_{j}$
is evident from this geometric picture.
In complete analogy with the usual Berry phase, this expression in
terms of a surface integral also proves the sensitivity of
$\gamma^\Gamma_{jk}$ to the presence of degeneracies of
the two energy level $i$ and $j$ in the parametric Hamiltonian
associated to the above paths.
However, given the open path $\Gamma$ and the energy levels involved, there
is no general rule to determine a closed loop in parameter's space
entangled with a degenerate submanifold.
Whenever this loop can be found, $\gamma^\Gamma_{jk}$ is
a direct probe of the presence and position of degeneracies.

The simplest system to illustrate the concept of off-diagonal geometric
phase is a spin-$\frac 12$ aligned to a slowly rotating magnetic field
$\bf B$ in (say) the $xz$ plane.
The polar angle $\theta$ of $\bf B$ 
parameterizes a circular path in the 2-dimen\-sional space of the magnetic
fields.
For any value of $\theta$, the
columns of the matrix
\begin{equation}
U(\theta)=
\pmatrix{	\cos\frac {\theta}2 &	\sin\frac {\theta}2\cr
		-\sin\frac {\theta}2  &	\cos\frac {\theta}2
		}
	\,.
	\label{umatrix:eq}
\end{equation}
represent the parallel-transported eigenvectors $|\psi_j(\theta)\rangle$ on
the initial basis $|\psi_1(0)\rangle=\left|\downarrow\right\rangle$,
$|\psi_2(0)\rangle=\left|\uparrow\right\rangle$.
Thus, the familiar Pancharatnam-Berry phase factor of the state
$|\psi_j(\theta)\rangle$ evolving from $\theta=0$ to $\theta_{\rm f}$ is
given by the diagonal matrix element $\gamma_j(\theta_{\rm f})=
\Phi\!\left( U_{jj}(\theta_{\rm f}) \right)
\Phi\!\left(\langle \psi_j(0) |\psi_j(\theta_{\rm f})\rangle \right)$.
The single off-diagonal term is
$\gamma_{12} = \Phi(\sin {\theta/2}) \,\Phi(-\sin \theta_{\rm f}/2) \equiv -1 $
for any $\theta_{\rm f} \neq 0$, $2\pi$.
For generic $\theta$, $\gamma_1$, $\gamma_2$ and $\gamma_{12}$ are all
equally important.
For $\theta=\pi$, $\gamma_{12}$ carries all the geometric phase
contents of the eigenstates, while $ \gamma_1$ and $\gamma_2$ are 
undefined.
At $\theta=2 \pi$ the roles are exchanged.
In this sense, the off-diagonal phase factor $\gamma_{12}$ constitutes the
counterpart of $\gamma_j$, when the latter is undefined.

Interference experiments \cite{Wagh98} have measured the noncyclic
Pancharatnam-Berry phases $\gamma_j$ in the spin-$\frac12$ system.
In a similar way, one can envisage a spin-rotation experiment to measure
by interference $\sigma_{12}$ and $\sigma_{21}$ for an arbitrary fixed
gauge at the starting point.
The dependence on the gauge chosen cancels out in the product
$\gamma_{12}$, which, for this simple system, must equal $-1$ for any
rotation angle $\theta\neq 2 \pi$.
Essentially any experiment\cite{Christian93,Weinfurter90,Wagh98}
sensitive to open-path diagonal geometric
phases can be generalized to observe off-diagonal phases.
In systems of larger dimensionality, several off-diagonal phase factors
can be defined, and they may assume different values on different paths.

The definition \refe{gammadef1:eq} of the off-diagonal phase factors
$\gamma^\Gamma$ can be generalized to the simultaneous evolution of more
than two orthonormal states.
Consider for example $n$ orthonormal eigenstates $|\psi_j(\sv)\rangle$
(ordered by increasing energy) of a parameterized Hermitian Hamiltonian
matrix $H(\sv)$, representing a physical system.
Observing the effect \refe{sigmagauge:eq} of a gauge change on the
$\sigma_{jk}^\Gamma$ phase factors, we note that {\em any cyclic product}
of $\sigma$'s is gauge-invariant.
It is then natural to generalize Eq.~\refe{gammadef1:eq} by defining
\begin{equation}
	\gamma_{j_1j_2j_3... j_l}^{(l)\,\Gamma}=
	\sigma_{j_1j_{2}}^\Gamma \,
	\sigma_{j_2j_{3}}^\Gamma \,
	\cdots \,
	\sigma_{j_{l-1}j_{l}}^\Gamma \,
	\sigma_{j_lj_{1}}^\Gamma \,
	\,.
	\label{gammadef2:eq}
\end{equation}
For $l=1$, Eq.~\refe{gammadef2:eq} reduces to the familiar definition
\refe{berrydef:eq} of the Pancharatnam-Berry diagonal phase factor
$\gamma_j^\Gamma=\gamma_{j}^{(1)\,\Gamma}=\sigma^\Gamma_{jj}$.
The 2-indexes $\gamma_{jk}^{(2)\,\Gamma}$ phase factors coincide with those
introduced by Eq.~\refe{gammadef1:eq}.  
Larger $l$ describe more complex phase relations among off-diagonal
components of the eigenstates at the endpoints of $\Gamma$.
The same geometrical construction of a closed path done for $\gamma^{(2)}$ 
extends to $\gamma^{(l)}$ with $l>2$.

We  note that any cyclic permutation of all the indexes $j_1j_2
j_3... j_l$ is immaterial.
Moreover, if one index is repeated, the associated $\gamma^{(l)}$ can be
decomposed into the product $\gamma^{(l_1)}\,\gamma^{(l_2)}$'s with
$l_1+l_2=l$.
We can thus reduce to consider the $\gamma^{(l)}$'s with no repeated
indexes, which means in particular $l\leq n$.

One can readily verify that the number of $\gamma^{(l)}$'s left grows with
$n$ faster than $n^2$.  Since $n^2$ is the number of the constituent
$\sigma_{jk}$'s, not all the $\gamma^{(l)}$'s can be independent.
We shall now find a complete set of independent $\gamma^{(l)}$'s, under the
condition that $U^{\Gamma}_{jk} \neq 0$ for all $j$ and $k$.
Clearly, the $n$ Pancharatnam-Berry diagonal phase factors $\gamma^{(1)}_j$
are all independent, since any diagonal $\sigma_{jj}$ enters only
$\gamma^{(1)}_j$.
On the other hand, the off-diagonal $\gamma^{(l)}$'s are interrelated by
the following exact equalities 
[they can be verified substituting explicitly the definition
\refe{gammadef2:eq}]:
\begin{eqnarray}
\gamma^{(l)}_{i \{j\}k\{m\}}&=&
	\gamma^{(l')}_{i \{j\}k}\gamma^{(l'')}_{k\{m\}i}
	{\gamma^{(2)}_{i k}}^*
	\ \ \ \ (l\geq 4)
\label{gammacomb23:eq}	\\
\gamma^{(3)}_{jkm} \gamma^{(3)}_{jmk}&=&
	\gamma^{(2)}_{jk}\gamma^{(2)}_{km}\gamma^{(2)}_{jm}
\label{gammacomb32:eq}	\\
\gamma^{(3)}_{ijm} {\gamma^{(2)}_{mj}}^* \gamma^{(3)}_{jkm}
&=&
\gamma^{(3)}_{ijk} {\gamma^{(2)}_{ki}}^* \gamma^{(3)}_{ikm}
\,.
\label{gammacomb4:eq}
\end{eqnarray}
In Eq.~\refe{gammacomb23:eq}, $\{j\}$ indicates a set of one or more
indexes, and $l'$, $l''$ ($<l$) count the indexes in the corresponding
$\gamma$.
Combining relations (\ref{gammacomb23:eq}-\ref{gammacomb4:eq}),
any $\gamma^{(l)}$'s may be expressed in terms of three
categories: the $n$ diagonal phases $\gamma^{(1)}_j$, the $n(n-1)/2$
quadratic $\gamma^{(2)}_{j<k}$'s, and the $(n-1)(n-2)/2$ cubic
$\gamma^{(3)}_{1<j<k}$.
These $n^2-n+1$ factors are indeed functionally independent combinations of
the $\sigma$'s: we verified that the Jacobian determinant $\left|\partial
\gamma_{\{j\}} / \partial \sigma_{km}\right|$ is nonzero.
The number of independent phases can be easily understood: it amounts 
to the $n^2$ phases of $U_{jk}^\Gamma$ minus the arbitrary
$n-1$ relative phases among the $n$ eigenstates at a given
point $\sv$.

We restrict now to the particular case of a path
joining a pair of points $\sv^P _1$ $\sv^P _2$ such that the $n$ 
eigenstates at the final point are a permutation $P$ of the initial 
eigenstates, {\em i.e.}
\begin{equation}
\left\{
\begin{array}{rcl}
  H(\sv^P _1)  & = & \sum_j E_j  | \psi_j \rangle \langle \psi_j |
\cr
  H(\sv^P _2)  & = & \sum_j E'_j  | \psi_{P_j} \rangle \langle \psi_{P_j} |
\cr
\end{array}
\right. \,,
\label{permut:eq}
\end{equation}
where $E_j$ and $E_j'$ are in increasing order as usual.
The only well-defined
$\sigma^\Gamma$'s are the $n$ phase factors $\sigma_{j\,P_j}^\Gamma$.
When the permutation is nontrivial ($P_j\neq j$) the familiar
Berry-Pancharatnam phase factor associated to state $j$ is
undefined. 
For this special case the only well-defined geometric phases 
are the off-diagonal ones. 
One can classify them according to standard group theory.
Any permutation $P$ can be decomposed univocally into $c$
cycles of lengths $l_1$, $l_2$, $\dots$ $l_c$ \cite{Hamermesh}.
To each cycle $i$, it is possible to associate one
$\gamma^{(l_i)\,\Gamma}_{\{j\}}$, the $l_i$ indexes $\{j\}$ following
the corresponding cycle. These phase factors involve only nonzero
$U^\Gamma_{jk}$ and are thus well defined.  In  contrast,
all other $\gamma^{(l)\,\Gamma}$'s are undefined. 
In Table~\ref{gammas:tab}, for each permutation $P$ of the eigenstates we
report the corresponding well-defined $\gamma^{(l)}$ for 
$n\leq 4$.


For these paths permuting the eigenvectors, the determinant
$\left|U^{\Gamma}\right|$ of the overlap matrix is related to the product
of the $\sigma$'s.
The equality $\left|U^\Gamma \right|=1$ becomes therefore
\begin{equation}
	\prod_{j=1}^n \sigma_{j\,P_j}^\Gamma = (-1)^P
	\,.
\label{detone:eq}
\end{equation}
The third column of Table~\ref{gammas:tab} summarizes this condition in
terms of the $\gamma^{(l)}$'s.
In the special case of a {\em real symmetric} Hamiltonian $H(\sv)$, all
$\sigma$'s, and thus all $\gamma^{(l)}$'s either equal $+1$ or $-1$.
For this simple but relevant situation, the last column of
Table~\ref{gammas:tab} reports the number of combinations of values that
the $\gamma^{(l)}$'s may take, as allowed by the condition
\refe{detone:eq}.

\vbox{
\begin{table}[b]
\begin{center}
\begin{tabular}{l|llll}
	&  		& geometric		& condition	&	\# of\\
$n$	& $P$ 		& phase factors  	& $\left|U^\Gamma \right|=1$
								&	cases\\
\hline
1	& 1 		& $\gamma_1$  	& $\gamma_1=1$		& 1 \\
\hline
2	& 1 2 		& $\gamma_1$ $\gamma_2$  	
					& $\gamma_1\,\gamma_2=1$	& 2 \\
	& 2 1 *		& $\gamma_{12}$	& $\gamma_{12}=-1$	& 1 \\
\hline
3	& 1 2 3		& $\gamma_1$ $\gamma_2$ $\gamma_3$
				& $\gamma_1\,\gamma_2\,\gamma_3=1$	& 4 \\
	& 2 1 3		& $\gamma_{12}$ $\gamma_3$
					& $\gamma_{12}\,\gamma_3=-1$	& 2 \\
	& 3 2 1 *	& $\gamma_{13}$ $\gamma_2$
					& $\gamma_{13}\,\gamma_2=-1$	& 2 \\
	& 1 3 2		& $\gamma_{23}$ $\gamma_1$
					& $\gamma_{23}\,\gamma_1=-1$	& 2 \\
	& 2 3 1		& $\gamma_{123}$	& $\gamma_{123}=1$	& 1 \\
	& 3 1 2		& $\gamma_{132}$	& $\gamma_{132}=1$	& 1 \\
\hline
4	& 1 2 3	4	& $\gamma_1$ $\gamma_2$ $\gamma_3$  $\gamma_4$
			& $\gamma_1\,\gamma_2\,\gamma_3\,\gamma_4=1$	& 8 \\
	& 2 1 3	4	   & $\gamma_{12}$ $\gamma_3$ $\gamma_4$
				& $\gamma_{12}\,\gamma_3\,\gamma_4=-1$	& 4 \\
	& [5 similar]	& ...	&	&	4\\
	& 4 3 2	1 *	   & $\gamma_{12}$ $\gamma_{34}$
				& $\gamma_{12}\,\gamma_{34}=1$	& 2 \\
	& [2 similar]	& ...	&	&	2\\
	& 2 3 1 4	& $\gamma_{123}$ $\gamma_4$
				& $\gamma_{123}\,\gamma_4=-1$	& 2 \\
	& [7 similar]	& ...	&	&	2\\
	& 2 3 4 1	& $\gamma_{1234}$  & $\gamma_{1234}=1$	& 1 \\
	& [5 similar]	& ...	&	&	1\\
\end{tabular}
\end{center}
\caption{
All possible geometric phase factors $\gamma^{(l)}$ defined in
Eq.~(\protect\ref{gammadef2:eq}), for an arbitrary path joining a point
$\sv _1$ to $\sv _2$, such that the eigenvectors of $H(\sv _2)$ are
permuted according to $P$ with respect to those of $H(\sv _1)$.  The last
column lists the number of the possible combinations of values ($\pm 1$)
that the $\gamma^{(l)}$ factors can take in the special case of a real
$H(\sv )$.  The stars mark the permutations induced by relation
(\protect\ref{symm:eq}), observed at the half-loop of
Ref.~\protect\cite{Lauber94} for $n=2$ and $3$.
\label{gammas:tab}
}
\end{table}
}

The above arguments on the permutational symmetry remain valid
even if Eq.~\refe{permut:eq} is only approximate, provided that
$|U_{j,P_j}^{\Gamma}|\gg n\,\max_{(k\neq P_j)} |U_{jk}^{\Gamma}|$ for all $j$.
This extends the interest of the permutational case to a finite domain of
the parameters' space around the point where Eq.~\refe{permut:eq} holds 
exactly  or, more in general, to any region where the
inequality on $U_{jk}^{\Gamma}$ holds.
For example, an approximate permutation occurs when the energy levels of an
Hamiltonian $H(\sv)$ undergo a sequence of sharp avoided crossings along
the path.
At each avoided crossing, the two involved eigenstates,
to a good approximation, exchange.
As a result, there exist sizable regions between two avoided crossings
where the eigenvectors are an approximate permutation of the starting ones.

Probably the simplest example of a nontrivial permutation of the
Hamiltonian eigenstates occurs when the relation
\begin{equation}
        H({\sv _1})=-H({\sv _2})
        \label{symm:eq}
\end{equation}
holds at the ends of the path.
This symmetry is verified exactly by the spin-$\frac 12$ system, where it
determines the swap of the eigenstates between $\theta=0$ and $\theta=\pi$.
Relation \refe{symm:eq} holds also, approximately, in very common
situations.
Suppose, for example, that a point, say $\sv =0$, locates an $n$-fold
degeneracy, and consider the perturbative expansion around there:
\begin{equation}
	H(\sv) 
	= 
	\sv \cdot {\bf H}^{(1)} +  \dots 
	\quad.
	\label{expansion:eq}
\end{equation}
[${\bf H}^{(1)}$ is a vector of Hermitian numerical matrices.]
In the sufficiently small neighborhood of the degeneracy, where the linear
term accounts for the main contribution to the energy shifts, pairs of
opposite points $({\sv _1},{\sv _2}=-{\sv _1})$ satisfy the relation
\refe{symm:eq}.
The permutation of the eigenstates associated to \refe{symm:eq} is composed
by $n/2$ 2-cycles for even $n$, or by $(n-1)/2$ 2-cycles plus one 1-cycle
for odd $n$: the corresponding $\gamma$'s are marked by stars
in Table~\ref{gammas:tab}.

\begin{figure}[tbh]
\centerline{ 
\psfig{file=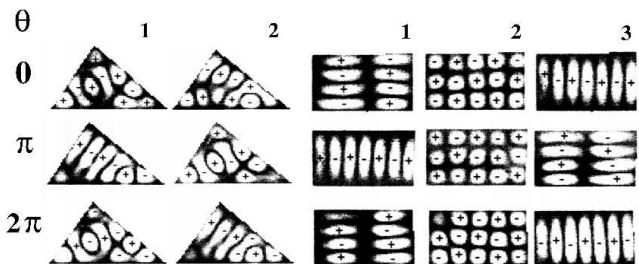,width=8.6cm}
}
\caption{
The observed initial ($\theta=0$), intermediate ($\theta=\pi$) and final
($\theta=2\pi$) eigenstates of the microwave cavities deformed following
adiabatically the path of Ref.~\protect\cite{Lauber94}.
Left: the two eigenstates of the triangular resonator.
Right: the three eigenstates of the rectangular resonator.
\label{expwaves:fig}
 }
\end{figure}

In the final part of this Letter, we examine the deformed
microwave resonators experiment of Ref.~\cite{Lauber94}.
In a recent work \cite{manolopoulos} the diagonal, closed-path
Berry phases were calculated for that system.
Here we analyze the experiment of Ref.~\cite{Lauber94} as a
transparent example of how off-diagonal $\gamma^{(2)}_{jk}$'s
can be measured for open paths.
For these systems, $\sv = (s \cos \theta, s \sin \theta)$ parameterizes
the displacement of one corner of the resonator away from the position of a
conical intersection of the energy levels.
Lauber {\it et al.} \cite{Lauber94} investigate the Berry phase of these
nearly degenerate states, when the distortion is driven through a loop
$\theta=0$ to $2 \pi$ around the degenerate point.
The distortion path is traced in small steps in $\theta$, following
adiabatically the real eigenfunctions.
In Fig.~\ref{expwaves:fig} we report the initial ($\theta=0$), half-way
($\theta=\pi$) and final ($\theta=2\pi$) parallel-transported
eigenfunctions from the original pictures of Ref.~\cite{Lauber94}.

The first case considered is that of a triangular cavity deformed around a
twofold degeneracy: for small distortions, the system behaves similarly to
a spin $\frac 12$.
In particular, the Berry phases $\gamma^{(1)}_j$ at the end of the loop
both equal $-1$ as expected for such a situation (cf.\ in
Fig.~\ref{expwaves:fig} the recurrence of the pattern with changed sign at
$\theta=0$ and $2\pi$).
Due to the well approximate symmetry \refe{symm:eq} at half path
($\theta=\pi$), the diagonal Berry phases are undefined there, but it is
instead possible to determine the experimental value of $\gamma^{(2)}_{12}$
for this path.
 From inspection of Fig.~\ref{expwaves:fig} we determine $\sigma_{12}=1$,
$\sigma_{21}=-1$.  This is consistent with the only possible value
$\gamma^{(2)}_{12}=-1$ allowed in this spin-$\frac 12$--like case (see Table
\ref{gammas:tab}).
The same holds for the path going from $\theta=\pi$ to $2\pi$.

The case of the rectangular resonator is more interesting.  Here, three
states intersect conically at $\sv=0$.
The three Berry phases $\gamma^{(1)}_j$ at the end of the loop ($-1$, $+1$
and $-1$) are compatible with the determinant requirement of Table
\ref{gammas:tab}.
Figure~\ref{expwaves:fig} shows that empirically also this system satisfies
the symmetry relation $H(\pi)=-H(0)$ at mid loop.
Thus, for the path $\theta=0$ to $\pi$ the only well defined
Pancharatnam-Berry phase is that of the central state $\gamma^{(1)}_2=-1$.
The upper and lower states exchange, giving $\sigma_{13}=1$,
$\sigma_{31}=1$ thus $\gamma^{(2)}_{13}=1$.
This is one of the two combinations of values allowed by the determinant
rule $\gamma_{13}\,\gamma_2=-1$ of Table \ref{gammas:tab}.

In conclusion, we have identified novel off-diagonal geometric phase
factors, generalizing the (diagonal) Berry phase.
The two sets of diagonal and off-diagonal geometric phases together
exhaust the number of independent
observable phase relations among $n$ orthogonal
states evolved along a path.
We show that, in many common situations, the off-diagonal factors carry the
relevant geometric phase information on the basis of eigenstates.

We thank Prof.\ D.\ Dubbers, Dr.\ F.\ Faure, and Dr.\ A.~F.\ Morpurgo for
useful discussions.

\bibliographystyle{prsty}


\begin{thebibliography}{10}
\bibitem{simon:note}
{ The Bargmann invariants of Ref.~\protect\cite{Simon93,Mukunda93,Rabei99} are
  undefined for the orthogonal states we consider in
  Eq.~(\protect\ref{gammadef2:eq})}.
\end{thebibliography}

\end{document}